\documentclass[12pt]{article}
\usepackage{hyperref}
\catcode`\@=11
\newcount\hour
\newcount\minute
\newtoks\amorpm
\hour=\time\divide\hour by60
\minute=\time{\multiply\hour by60 \global\advance\minute by-\hour}
\edef\standardtime{{\ifnum\hour<12 \global\amorpm={am}%
        \else\global\amorpm={pm}\advance\hour by-12 \fi
        \ifnum\hour=0 \hour=12 \fi
        \number\hour:\ifnum\minute<10 0\fi\number\minute\the\amorpm}}
\edef\militarytime{\number\hour:\ifnum\minute<10 0\fi\number\minute}
\def\draftlabel#1{{\@bsphack\if@filesw {\let\thepage\relax
   \xdef\@gtempa{\write\@auxout{\string
      \newlabel{#1}{{\@currentlabel}{\thepage}}}}}\@gtempa
   \if@nobreak \ifvmode\nobreak\fi\fi\fi\@esphack}
        \gdef\@eqnlabel{#1}}
\def\@eqnlabel{}
\def\@vacuum{}
\def\draftmarginnote#1{\marginpar{\raggedright\scriptsize\tt#1}}
\def\draft{\oddsidemargin -.2truein
        \def\@oddfoot{\sl preliminary draft \hfil
        \rm\thepage\hfil\sl\today\quad\militarytime}
        \let\@evenfoot\@oddfoot \overfullrule 3pt
        \let\label=\draftlabel
        \let\marginnote=\draftmarginnote
   \def\@eqnnum{(\theequation)\rlap{\kern\marginparsep\tt\@eqnlabel}%
\global\let\@eqnlabel\@vacuum}  }
\relax
\def\sqr#1#2{{\vcenter{\vbox{\hrule height.#2pt
        \hbox{\vrule width.#2pt height#1pt \kern#1pt
           \vrule width.#2pt}
        \hrule height.#2pt}}}}

\def\lsim{{\displaystyle
{{\raise-8pt\hbox{$ <$}}
\atop{\raise5pt\hbox{$\sim$}}}}}
\def\gsim{{\displaystyle
{{\raise-8pt\hbox{$ >$}}
\atop{\raise5pt\hbox{$\sim$}}}}}
\def\slsim{{\displaystyle
{{\raise-8pt\hbox{$\scriptstyle <$}}
\atop{\raise5pt\hbox{$\scriptstyle \sim$}}}}}
\def\sgsim{{\displaystyle
{{\raise-8pt\hbox{$\scriptstyle  >$}}
\atop{\raise5pt\hbox{$\scriptstyle \sim$}}}}}
\newcommand{\ar}[2]{{#1\atopwithdelims[]#2}}
\newcommand{\sump}[0]{\sum_{(h,g)}\!{\raise 4pt \hbox{$'$}}\,}
\def\Zint{\mbox{\sf Z\hspace{-3.2mm} Z}}
\def\Zints{\rm Z}
\def\Real{\mbox{I\hspace{-2.2mm} R}}

\def\pa{\partial}
\def\bpa{\bar{\partial}}

\def\ra{\to}

\def\ti{\times}
\def\ss{\ar{a}{b}}
\def\bss{\ar{\ba}{\bb}}

\def\w{\wedge}
\def\Tr{\,{\rm Tr}\, }
\def\det{\,{\rm det}\, }

\def\Im{\,{\rm Im}\, }
\def\Re{\,{\rm Re}\, }

\def\a{\alpha}
\def\b{\beta}

\def\e{\epsilon}
\def\m{\mu}
\def\n{\nu}
\def\t{\tau}
\def\p{\pi}
\def\ps{\psi}

\def\r{\rho}
\def\th{\vartheta}

\def\s{\sigma}
\def\l{\lambda}
\def\k{\kappa}

\def\f{\phi}

\def\Ga{\Gamma}

\def\z{\zeta}
\def\bal{\bar{\alpha}}
\def\bbe{\bar{\beta}}

\def\bz{\bar{z}}

\def\ba{\bar{a}}
\def\bb{\bar{b}}

\def\bps{\bar{\psi}}

\def\bT{\overline{T}}
\def\bU{\overline{U}}

\def\Z{Z\!\!\! Z}

\def\F{{\cal F}}

\def\T{{\cal T}}

\def\sp{\; , \; \; }

\def\rd{{\rm d}}

\def\ed{\end{document}}
\newtoks\@stequation
\def\subequations{\refstepcounter{equation}%
  \edef\@savedequation{\the\c@equation}%
  \@stequation=\expandafter{\theequation}
  \edef\@savedtheequation{\the\@stequation}
  \edef\oldtheequation{\theequation}%
  \setcounter{equation}{0}%
  \def\theequation{\oldtheequation\alph{equation}}}

\def\endsubequations{\setcounter{equation}{\@savedequation}%
  \@stequation=\expandafter{\@savedtheequation}%
  \edef\theequation{\the\@stequation}\global\@ignoretrue
  \vspace*{-12pt} \\}

\def\be{\begin{equation}}
\def\ee{\end{equation}}
\def\bs{\begin{subequations}}
\def\es{\end{subequations}}
\newskip\humongous \humongous=0pt plus 1000pt minus 1000pt
\def\caja{\mathsurround=0pt}
\def\eqalign#1{\,\vcenter{\openup1\jot \caja
        \ialign{\strut \hfil$\displaystyle{##}$&$
        \displaystyle{{}##}$\hfil\crcr#1\crcr}}\,}
\newif\ifdtup

\def\thebibliography#1{%
\vskip 0.5cm \centerline{\bf References}
\list{%
[\arabic{enumi}]}{\settowidth\labelwidth{[#1]}
\leftmargin\labelwidth
\advance\leftmargin\labelsep
\usecounter{enumi}}
\def\newblock{\hskip .11em plus .33em minus .07em}
\sloppy\clubpenalty4000\widowpenalty4000
\sfcode`\.=1000\relax}

\renewcommand{\theequation}{\arabic{section}.\arabic{equation}}

\renewcommand{\section}{\setcounter{equation}{0}\@startsection%
{section}{1}{0mm}{-\baselineskip}{0.5\baselineskip}%
{\normalfont\large\bfseries}}

\renewcommand{\subsection}{\@startsection%
{subsection}{2}{0mm}{-\baselineskip}{0.5\baselineskip}%
{\normalfont\normalsize\itshape}}

\begin{document}

\begin{titlepage}
\begin{flushright}
CERN-TH/97-146\\
CPTH-S539.0697\\
hep-th/9707018v6
\end{flushright}
\vspace{.1in}

\begin{center}
{\boldmath \large \bf On $R^4$ threshold corrections in type IIB
string theory \\
and $(p,q)$-string instantons }
\vspace{.5in}

{\bf Elias Kiritsis and Boris Pioline$^\diamond$}\\
\vspace{.2in}

{\em Theory Division, CERN$^\dagger$,\\ CH-1211,
Geneva 23, SWITZERLAND} \\

\vspace{1in}

\end{center}

\vskip .15in

\begin{center} {\bf ABSTRACT }
 \end{center}
\begin{quotation}\noindent
We obtain the exact non-perturbative thresholds of $R^4$ terms in
type IIB
string theory compactified to eight and seven dimensions.
These thresholds are given by the perturbative tree-level and one-loop
results together with the contribution of the D-instantons and
of the $(p,q)$-string instantons. The invariance under $U$-duality
is made manifest by rewriting the sum as a non-holomorphic
invariant modular function of the corresponding discrete $U$-duality group.
In the eight-dimensional case, the threshold
is the sum of a order-1 Eisenstein series for $SL(2,\Zint)$
and a order-3/2 Eisenstein series for $SL(3,\Zint)$.
The seven-dimensional result is given by the
order-3/2 Eisenstein series for $SL(5,\Zint)$.
We also conjecture formulae for the non-perturbative thresholds in
lower dimensional compactifications and discuss the relation
with M-theory.

\end{quotation}
\vskip 2cm

\begin{flushleft}
CERN-TH/97-146\\
CPTH-S539.0697\\
July 1997\\
\end{flushleft}
\hrule width 6.7cm \vskip.1mm{\small \small \small
$^\diamond$ On leave from { Centre de Physique Th{\'e}orique,
Ecole Polytechnique,} \\
{ CNRS,} 91128 Palaiseau Cedex, France.\\
$^\dagger$ e-mail addresses: kiritsis,bpioline@mail.cern.ch.}
\end{titlepage}
\vfill
\eject
\def\baselinestretch{1.2}
\baselineskip 14 pt

\noindent
\section{Introduction}
\setcounter{equation}{0}

Important progress has recently been made towards understanding the
non-perturbative structure of string theory.
Extended supersymmetry implies non-perturb\-ative
equivalence between string theories. Quantitative tests, although scarce, 
can be carried out by computing threshold corrections
to specials terms in the effective action,
which are ``BPS saturated''. Such terms
receive perturbative corrections from (at most) a single order in
perturbation theory, to which only BPS states contribute.
Usually they are related by supersymmetry to anomaly cancelling terms.
Moreover, they receive instanton corrections from special instanton
configurations
that leave some of the supersymmetries unbroken.
Examples are D-terms in N=1 theories in four dimensions, the two
derivative
terms in N=1 six-dimensional theories, the $F^4$ and $R^4$ terms in
theories with N=1 ten-dimensional supersymmetry \cite{BK,BFKOV} and
$R^4$ terms in ten-dimensional N=2 theories \cite{gg}.

Thresholds of ``BPS saturated'' terms become more complicated as the
theory is compactified to lower dimension without breaking the
supersymmetry.
The reason is that the number of scalar moduli that they can depend
upon grows, and the duality symmetries become larger as one decreases
the number of non-compact dimensions.

Here we will mainly focus on the various $R^4$ terms in type IIB string
theory toroidally compactified to eight and seven dimensions.
In ten dimensions such terms have tree-level and one-loop corrections
that have been computed \cite{gg} and are believed to receive
no other perturbative corrections.
On the other hand, the type IIB string has instantons already in ten
dimensions, known as D-instantons.
Their contribution was conjectured in Ref. \cite{gg}
on the basis of the $SL(2,\Zint)$ symmetry of the type IIB theory.
Upon compactification on a circle, nothing exotic happens.
Since the D-instanton contributions are independent of the non-compact
dimensions the nine-dimensional thresholds can be obtained from the
ten-dimensional exact result and from the nine-dimensional
perturbation theory \cite{gg}.

Upon compactification to eight dimensions a new type of instanton
enters the
game, namely (Euclidean) $(p,q)$-strings
whose world-sheets wrap around the target space torus.
One of the  main points in this paper is to calculate their
contributions
from first principles and show that the full result is
$SL(3,\Zint)\times SL(2,\Zint)$
invariant (the $U$-duality group in eight dimensions).
This result also matches a
recent proposal for the same threshold calculated in the context of
M-theory
\cite{vg2}.

Compactifying type II string theory further down to seven dimensions,
we do not expect anything exotic to happen.
The only difference is that now the Euclidean $(p,q)$ instantons are
wrapping
in all possible ways on two dimensional submanifolds of the
three-torus.
This contribution can be evaluated from the perturbative
world-sheet instantons of the fundamental type IIB string on $T^3$.
This will enable us to derive an expression for the exact
$R^4$ threshold in seven dimensions, that exhibits manifest
$SL(5,\Zint)$ $U$-duality symmetry.

Generalizing the above pattern, we will also propose
an exact expression for the six-dimensional case.
This expression is manifestly invariant under the $SO(5,5,\Zint)$
$U$-duality group, and should reproduce the contributions of
the D-instantons and D$(p,q)$-strings, together with the
contribution of the D-3brane that can be wrapped on the four-torus.
Lower dimensional cases lead us into the realm of discrete exceptional
groups, for which we will have little to say here.

Our proposal for the threshold is as follows. Let $G/H$ be the
homogeneous
space describing the scalars of a given compactification.
When this coset space is irreducible, the kinetic terms of the scalars
can be written in terms of the matrix $M$ in the adjoint representation
of
G
as
\be
S_{\rm  scalars}=\int d^Dx~ tr(\pa M\pa M^{-1})
\ee
Then, we conjecture that the threshold will be given by the order-3/2
Eisenstein series
\be
E_{3\over 2}(M)=\sum_{m^i} \left[\sum_{i,j}m^iM_{ij}m^j\right]^{-{3\over
2}}
\ee
The cases $D=9$ and $D=8$ have the peculiarity that the coset space
is reducible. The nine-dimensional case was investigated in Ref. \cite{vg1}.
In eight dimensions, which we analyze in detail here, the scalar manifold
splits into $G/H=SL(3,\Real)/SO(3)\times SL(2,\Real)/SO(2)$.
In this case the threshold will be given
by the sum of a order-3/2 $SL(3,\Zint)$
and a order-1 $SL(2,\Zint)$ series.
The seven-dimensional threshold will be shown
to be given by a order-3/2 $SL(5,\Zint)$ series.

The structure of this paper is as follows.
In Section 2 we briefly review the situation in ten dimensions.
In Section 3 we compactify to eight dimensions, calculate the
perturbative contributions corresponding to the
fundamental $(1,0)$ string, and then
generalize to $(p,q)$-strings.
In Section 4 we make the invariance the result in Section 3
under $U$-duality manifest, and show the presence of a logarithmic
correction overlooked by our argument in Section 3.
In Section 5 we  compactify to lower
dimensions, calculate the seven-dimensional threshold and consider
briefly (and incompletely) the six-dimensional case.
Finally, Section 6 contains our conclusions.
In Appendix A we present some useful formulae on the expansion and
regularization of $SL(3,\Zint)$ Eisenstein series.

\section{The IIB string in ten dimensions}

The lowest-order bosonic effective action of the type IIB string in ten
 dimensions in the Einstein frame is \cite{sch}
\be
S_{10}^E={1\over 2\kappa_{10}^2}\int d^{10}x\sqrt{-g_E}\left[ R
-{1\over
2}{\pa\tau\pa\bar\tau\over \tau_2^2}-{1\over 12\tau_2}(\tau
H_{N}+H_R)_{\mu\nu\rho}(\bar\tau H_N+H_R)^{\mu\nu\rho}\right]
\label{1}\ee
with $\kappa_{10}^2=2^6\cdot \pi^7\alpha'^4$\footnote{From now on we
set
$\alpha'=1$.}.
We have set the self-dual four-form to zero since it will not play any
role in the subsequent discussion.
The complex scalar $\tau$ contains the dilaton (string coupling
constant) as well as the Ramond-Ramond axion:
\be
\tau=a+ie^{-\phi}
\label{2}\ee
while as usual
\be
H^\a_{\mu\nu\rho}=\pa_{\mu}B^\a_{\nu\rho}+{\rm cyclic}
\label{3}\ee
is the field strength of the R-R ($\a=1$ or $R$)
or NS-NS ($\a=2$ or $N$) antisymmetric tensor.
Transforming to the string $\sigma$-model frame
$g_E=e^{-\phi/2}{g_\sigma}$ we obtain
\be
S_{10}^{\sigma}={1\over 2\kappa_{10}^2}\int
d^{10}x\sqrt{-g_{\sigma}}\left[e^{-2\phi}\left(
R+4(\pa\phi)^2-{1\over 12}H_N^2\right)-{1\over 2}(\pa a)^2-{1\over 12}
(aH_N+H_R)^2\right]
\label{6}\ee
The NS-NS fields have the usual tree-level dilaton dependence (the
string coupling is $g=e^{\phi}$), while the R-R couplings have no dilaton
dependence at tree
level.

The effective action is invariant under an $SL(2,\Real)_\tau$ symmetry
\be
\tau\to {a\tau+b\over c\tau+d}\;\;\;,\;\;\;ad-bc=1
\label{4}\ee
\be
\left(\matrix{B_N\cr B_R\cr}\right)\to \left(\matrix{d& -c\cr -b&
a\cr}\right)
 \left(\matrix{B_N\cr B_R\cr}\right)
\label{5}\ee
while the Einstein metric and the self-dual four-form are invariant.
A discrete subgroup of this symmetry, $SL(2,\Zint)_\tau$,
is conjectured to be an
exact non-perturbative symmetry of the IIB string. It acts as $PSL(2,\Zint)$
on the complex scalar
$\tau$ plus the charge-conjugation symmetry $\tau\to \tau$, $B^\a\to
-B^\a$.
Various arguments for this symmetry have been given \cite{sl},
including
the construction of the D1 and D3 branes as well as all the $(p,q)$
strings.

Because of the large supersymmetry, the leading terms in the effective
action
that have non-zero quantum corrections are terms with eight derivatives
including $t_8t_8R^4$, $\epsilon_{10}\e_{10}R^4$ and $\e_{10}t_8 B_N
R^4$. The tensor $t_8$ is the standard eight-index tensor arising
in string amplitudes,
while $\e_{10}$ is the
ten-dimensional Levi-Civita tensor\footnote{We use the same
normalization as in
\cite{gsw,tsey}, namely
$t_8 F^4 := t^{\a_1\a_2\dots\a_8}
F_{\a_1\a_2} \dots F_{\a_7\a_8} = 24 F^4 - 6 (F^2)^2$
for any antisymmetric matrix $F$,
omitting the Levi-Civita tensor of \cite{gsw}.
In particular, $\e_{10}t_8 R^4 = 24 \Tr R^4 - 6 (\Tr R^2)^2$
and $t_8 t_8 R^4 = 24 t_8 \Tr R^4 - 6 t_8 \Tr R^2 \Tr R^2
=12 (R_{\m\n\r\s}R^{\m\n\r\s})^2+\dots$
The expression
$\e_{10} \e_{10} R^4=-96 (R_{\m\n\r\s}R^{\m\n\r\s})^2+\dots$ is
the continuation to ten dimensions of twice the eight-dimensional Euler
density $2\e_8 \e_8 R^4$. We take the spacetime to be minkowskian.
}.
In the case of N=1 ten-dimensional supersymmetry
it was shown that these terms must combine into two different
superinvariants \cite{Roo} whose bosonic parts are
\be
J_0 = t_8 t_8  R^4 + {1 \over 8} \e_{10} \e_{10} R^4
\sp
J_1 = t_8 t_8  R^4 - {1\over 4} \e_{10} t_8 B_N R^4
\ee
We shall assume that this still holds for the case
of N=2 supersymmetry in ten dimensions, and this is certainly
in agreement with our results.
In particular, $J_1$ contains a CP-odd anomaly related coupling
and is therefore believed to receive only one-loop corrections.
The  $J_0$ invariant is expected  not to receive perturbative
corrections
beyond one loop but it is not protected from
non-perturbative corrections.
Indeed it will be
the purpose of this paper to obtain an exact non-perturbative
result for this coupling.

We will now describe the $R^4$ couplings in the two ten-dimensional
type II string theories in the hope of clarifying some confusion in
the literature\footnote{We would like to thank A. Tseytlin for sharing
his insights on the issue.}.
At tree level the effective action for these terms has been calculated
from S-matrix and $\s$-model computations \cite{gri} to be:
\be
S^{\rm tree}_{R^4}=2\z (3){\cal N}_{10}\int d^{10}x\sqrt{-g_{\sigma}}
{}~\tau_2^2~\left(t_8t_8+{1\over 8}\e_{10}\e_{10}\right)R^4
\label{a1}\ee
$$
=2\z (3){\cal N}_{10}\int d^{10}x\sqrt{-g_{E}}
{}~\tau_2^{3\over 2}~\left(t_8t_8+{1\over 8}\e_{10}\e_{10}\right)R^4
$$
in the $\sigma$ and Einstein frames, respectively.
The coefficient is given by
\be
{\cal N}_{10}={1\over 3\cdot 2^8\kappa_{10}^2}
\label{a2}\ee
This result holds both for type IIA and IIB ten-dimensional
superstrings,
and does not involve the  CP-odd $J_1$ coupling.
Since in the type II string theory supersymmetry is respected by the loop
expansion,
this confirms the statement that $J_0$ is still a superinvariant for
$N=2$ ten-dimensional supersymmetry.

At one-loop one finds a non-zero CP-even contribution
\be
S^{\rm 1-loop}_{R^4}={2\pi^2\over 3}{\cal N}_{10}\int
d^{10}x\sqrt{-g_{\sigma}}
{}~\left(t_8t_8 \pm {1\over 8}\e_{10}\e_{10}\right)R^4
\label{a3}
\ee
where the $+$ sign (resp. $-$) occurs for IIB (resp. IIA) theories.
There is furthermore a one-loop contribution to the CP-odd
$\e_{10}t_{8} R^4$ term in the type IIA theory.
Both these results can be inferred from our eight-dimensional
calculation
in section 3.2 but we have checked them also directly in ten
dimensions.
The one-loop
IIB threshold therefore reduces to the $J_0$ invariant,
while the type IIA threshold involves a combination
$2 J_1 -J_0$ of the two invariants.
This is compatible
with the D=11 supergravity limit of type IIA, while the vanishing
of $\e_{10}t_8 R^4$ coupling is compatible with the
type IIB $SL(2,\Zint)_{\tau}$ symmetry.
Further consistency checks are obtained under
compactification of IIA/B superstring on $K3$. The
reduction of $J_0$ on $K3$ is zero, so that Eq.\ (\ref{a2}) implies
the absence of tree-level $R^2$ coupling in $N=4$ type II vacua.
Eq.\ (\ref{a3}) further implies that there are no one-loop $R^2$
corrections in six-dimensional IIB/$K3$ superstring, while
the $R^2$ threshold is equal to the Euler number of $K3$ in
the type IIA case. Moreover one-loop $B\w R\w R$ terms have to occur
only in six-dimensional type IIA theory, as found by an explicit
calculation \cite{sixguys}.

It is strongly suspected that there are no further perturbative
contributions to these terms.
On the other hand, in the type IIB theory there can be non-perturbative
contributions due to the D-instantons \cite{gg}. Indeed,
the perturbative result is not invariant under the $SL(2,\Zint)_{\tau}$
symmetry.
In \cite{gg} the exact non-perturbative threshold for the $t_8t_8R^4$
term (or more accurately for the $J_0$ superinvariant)
was conjectured, by covariantizing the perturbative result under
the $SL(2,\Zint)_\t$ symmetry:
\be
S_{R^4}= {\cal N}_{10}\int d^{10}x\sqrt{-g_{E}}
{}~f_{10}(\tau,\bar\tau)~(t_8t_8~R^4 + {1\over 8} \e_{10}\e_{10} R^4)
\label{a4}\ee
where
\be
f_{10}(\tau,\bar\tau)=\hat{\sum_{m,n\in \Zints}}{\tau_2^{3/2}\over
|m+n\tau|^3}
\label{a5}\ee
A hat over a sum indicates omission of the (0,0) term.
$f_{10}$ is manifestly $SL(2,\Zint)$-invariant.
It can be expanded at large $\tau_2$ to reveal the perturbative and
D-instanton contributions
\be
f_{10}=2\z(3)\tau_2^{3/2}+{2\pi^2\over 3}{1\over \sqrt{\tau_2}}+
8\pi\sqrt{\tau_2}\sum_{m\not=0}\sum_{n=1}^{\infty}e^{2\p i mn\tau_1}
\left|{m\over n}\right|K_{1}(2\pi |m|n\tau_2)
\label{a6}\ee
$K_1$ is the $K$ Bessel function defined in the Appendix.
The first term is the tree-level term, the second one corresponds to
the one-loop correction,
while the rest are exponentially suppressed
as $e^{-1/g}$ at weak coupling and come
from D-instantons.
The threshold for a circle compactification
of type IIB was further obtained in \cite{vg1}.
The one for the type IIA theory follows from
$R\to 1/R$ duality.

We finally mention that the results above imply the
following terms in the D=11 M-theory effective action
\be
\delta S_{11}\sim \int d^{11}x\left[\sqrt{-g}(t_8t_8-
{1\over 4\cdot 3!}\epsilon_{11}\epsilon_{11})R^4-
{1 \over 4}\epsilon_{11} t_8 C R^4
\right]
\ee
where $C$ is now the three-form gauge potential of D=11 supergravity.
Upon compactification to ten dimensions this reproduces the
one-loop terms of the IIA string.
The tree level term cannot be seen in the eleven-dimensional
limit. The reason is that it is produced in the ten-dimensional theory
by integrating out the massive modes of the perturbative IIA string.
In the decompactification limit, these masses become infinite
and this term disappears.

\section{The type IIB string in eight dimensions}
\setcounter{equation}{0}

\subsection{\bf Scalar manifold and U duality}
We will now compactify the type IIB string on a two-torus. The
effective
tree-level action can be obtained by the standard Kaluza-Klein
reduction.
The scalar fields in eight dimensions are, in addition to the
complex scalar
$\tau$,
the two-dimensional $\sigma$-frame metric of the two-torus
\be
G_{IJ}={V\over U_2}\left(\matrix{1&U_1\cr
U_1&|U|^2\cr}\right)
\label{7}\ee
written in terms of the volume $V=\sqrt{G}$ and of the complex structure
modulus U of the two-torus, together with the two scalars coming from
the two
antisymmetric tensor backgrounds:
\be
B^\alpha_{IJ}=\epsilon_{IJ}B^\alpha
\label{8}\ee
The (physical) volume of the two-torus is $(2\pi)^2 V$.
In eight dimensions, in addition to the ten-dimensional
$SL(2,\Zint)_{\tau}$
symmetry
we also expect the usual T-duality symmetry $SL(2,\Zint)_T$
acting on the T-modulus
\be
T=B_N+i V
\label{888}\ee
as well as $SL(2,\Zint)_U$ acting on the U modulus.
Moreover the exchange T$\leftrightarrow$U maps the IIB  to the
type IIA string.

We omit the calculational details and present the Einstein-Hilbert action and
scalar kinetic terms in the eight-dimensional Einstein frame:
\be
S_8^{E}={1\over 2 \k_8^2} 
\int d^8x\sqrt{-g_E}\left[R-{1\over 6}{\pa\n^2\over \n^2}
-{1\over 2}{\pa U\pa\bar U\over U_2^2}-{1\over
2}{\pa\tau\pa\bar\tau\over
\tau_2^2}
-\n{|\tau\pa B_N+\pa B_R|^2\over 2\tau_2}+\cdots\right]
\label{9}\ee
where $\k_8=\k_{10}/2\pi$.The scalar
\be
\n={1\over \tau_2 V^2}
\label{10}\ee
is invariant under the $SL(2,\Zint)_{\tau}$ duality.
The above action is manifestly invariant under the
$SL(2,\Zint)_{\tau}\times SL(2,\Zint)_U$ part of the duality.
The $SL(2,\Zint)_T$ duality becomes manifest
(and $SL(2,\Zint)_\tau$ duality hidden) if we introduce the $T$
variable
(\ref{888}) together with the T-duality invariant eight-dimensional
dilaton
\be
e^{\l}={1\over V\tau_2^2}
\label{11}\ee
which is the standard eight-dimensional string expansion parameter, and
with the complex scalar
\be
\xi=-B_R+ia V
\label{12}\ee
The action can now be written as
\be
S_8^{E}={1\over 2\k_8^2}
\int d^8x\sqrt{-g_E}\left[R-{1\over 6}(\pa \l)^2-{1\over 2}{\pa
U\pa\bar U\over U_2^2}-{1\over 2}{\pa T\pa\bar T\over T_2^2} -
\right.
\label{13}
\ee
$$
\left.
-{e^{\l}\over 2 T_2}
\left|
\pa \left({{\rm Im}(T\bar \xi)\over T_2}\right)
+T \pa\left({{\rm Im}\xi\over T_2}\right) 
\right|^2 \right]
$$
Under an $SL(2,\Zint)_T$ transformation
\be
T\to {aT+b\over cT+d}\;\;\;,\;\;\;\xi\to {\xi\over cT+d}
\label{14}\ee
the rest being inert,
the effective action is indeed invariant.
The field $\xi$ transforms as the complex coordinate on a torus of
complex structure $T$.
The maximal invariance of the action in Eq.\ (\ref{9}) is known to be
larger, namely
$SL(3,\Real)\times SL(2,\Real)$ \cite{julia}.
Its discrete subgroup $SL(3,\Zint) \times SL(2,\Zint)_U$ was conjectured
\cite{ht}
to be an exact symmetry ($U$-duality) of the eight-dimensional IIB
string.
The symmetry becomes manifest if we introduce the following
symmetric matrix with determinant one:
\be
M=\n^{1/3}\left(\matrix{{1\over \tau_2}&{\tau_1\over
\tau_2}&{{\rm Re}(B)\over \tau_2}\cr {\tau_1\over
\tau_2}&{|\tau|^2\over
\tau_2}& {{\rm Re}(\bar \tau B)\over \tau_2}\cr  {{\rm Re}(B)\over
\tau_2}&{{\rm Re}(\bar \tau B)\over \tau_2}&
{1\over \n}+{|B|^2\over \tau_2}\cr
}\right)\;\;\;,\;\;\;M=M^T\;\;,\;\;
{\rm det}(M)=1
\label{15}\ee
where we introduced the complex scalar $B=B_R+\tau B_N$ that transforms
the same way
as $\xi$ under $SL(2,\Zint)_{\tau}$ duality.
The matrix $M$ parametrizes the $SL(3,\Real)/SO(3)$ coset. In terms of $M$
the effective action can be written as
\be
S_8^{E}={1\over 2\k_8^2}
\int d^8x\sqrt{-g_E}\left[R-{1\over 2}{\pa U\pa\bar U\over
U_2^2}
+{1\over 4}{\rm Tr}(\pa M\pa M^{-1})\right]
\label{16}
\ee
which shows that the scalar manifold is
$SL(3,\Real)/SO(3)\times SL(2,\Real)/SO(2)$. An element $\Omega$
of $SL(3,\Real)$ acts on $M$ as
$M\to \Omega M\Omega^T$.
The conjectured $SL(3,\Zint)$ part of the $U$-duality symmetry
is generated by matrices $\Omega$ with integer
entries.
This symmetry can be obtained by intertwining $SL(2,\Zint)_{\tau}$ and
$SL(2,\Zint)_T$ transformations, which are embedded in $SL(3,\Zint)$
as follows:
\be
{SL(2,\Zint)}_{\tau}\;\;:\;\;\left(\matrix{a&b&0\cr c&d&0\cr
0&0&1\cr}\right)
\;\;\;,\;\;\;{SL(2,\Zint)}_{T}\;\;:\;\;\left(\matrix{1&0&0\cr 0&a&b\cr
0&c&d\cr}\right)
\label{20}\ee

The $SL(3,\Zint)$ part of the $U$-duality
symmetry is easily understood in the dual IIA
theory.
This is M-theory compactified on a three-torus with metric $G_3$. The
volume
$\sqrt{G_3}$ together with the three-index antisymmetric tensor
$C_{123}$
corresponds to the type IIB modulus $U$.
The remaining metric with unit determinant $\tilde G_3={\rm
det}(G)^{-1/3}G_3$
corresponds to the IIB $SL(3,\Zint)$ matrix $M$, and
$SL(3,\Zint)$ is the modular group of the 3-torus.

\subsection{\bf Perturbative gravitational thresholds}

We will parametrize the eight-dimensional threshold corrections as
\be
S^{R^4}_8={\cal N}_{8}\int d^8 x\sqrt{-g_E}\left[
(\Delta_{tt}t_8t_8+{1\over 4}\Delta_{\e\e}\e_{8}\e_{8}+
\Theta~t_8\e_8 )R^4\right]
\label{b1}
\ee
where ${\cal N}_8=(2\pi)^2{\cal N}_{10}$ and we are in the
eight-dimensional Einstein frame (note that $\Delta_{tt},\Delta_{\e\e}$
and $\Theta$ are all dimensionless in eight dimensions).
The ten-dimensional result discussed in section 2 implies the following
large-volume behaviour of the thresholds
\be
\lim_{T_2\to \infty}{\Delta_{tt}\over T_2}=
\lim_{T_2\to \infty}{\Delta_{\e\e}\over T_2}=
\sqrt{\tau_2}~f_{10}(\tau,\bar\tau)
\sp
\lim_{T_2\to \infty}{\Theta\over T_2}=0
\label{b2}
\ee
The tree-level result is obtained directly by compactification
of the ten-dimensional result:
\be
\Delta_{tt}^{\rm tree}=2\z(3)V\tau_2^{2}=2\z(3){\tau_2^{3/2}\over
\n^{1/2}}
\label{b3}
\ee

The one-loop result can also be directly computed by evaluating
the one-loop scattering amplitude of four gravitons together with
one modulus field of $T^2$, using the same techniques as
for the four-dimensional $R^2$ terms
\cite{sixguys}. The one-loop $R^4$ thresholds are also IR-divergent,
and can be regularized by introducing an IR cutoff by hand.
\footnote{This regularization breaks modular invariance. A stringy IR
regularization method was developped in \cite{kk},
but in this case it breaks the supersymmetry. In the cases where
it can be applied, it agrees with a
the usual non-modular invariant regularization.}
This introduces an ambiguity in the moduli independent part
of the threshold.

We therefore compute the following amplitude:
\be
{\cal A} =
\int_{\cal F} {\rd^2 \t \over \t_2^2} \prod_{i=1}^4
\langle \int {\rd^2 z_i  \over \p}
\e^i_{\bal_i \a_i} V^{\bal_i \a_i}(p_i,\bz_i,z_i)
\int {\rd^2 z_5  \over \p}  V_\f(p_5,\bz_5,z_5)
\rangle
\label{64}
\ee
In this expression, $\e^i_{\a\b}$ denote the transverse
symmetric traceless polarization
tensors of the four gravitons, whose vertex operators in the zero ghost
picture
read
\be
V^{\bal \a} (p,\bz,z) =
[\bpa X^{\bal} (\bz,z) + i p \cdot \bps (\bz) \bps^{\bal} (\bz) ]
[\pa X^\a (\bz,z) + i p \cdot \ps (z) \ps^\a (z)]
{\rm e}^{i p \cdot X (\bz,z) }
\label{vop}
\ee
The modulus field vertex operator is defined in a similar way:
\be
\label{mvop}
V_\f(p,\bz,z) =
v_{IJ}(\f)  [\bpa X^I (\bz,z) + i p \cdot \bps (\bz) \bps^I (\bz)]
[ \pa X^J (\bz,z) + i p \cdot \ps (z) \ps^J (z) ]
{\rm e}^{i p \cdot X (\bz,z) }
\ee
in terms of the metric and antisymmetric fields in the internal $T^2$
directions:
\be
v_{IJ}(\f) = \pa_\f (G_{IJ} + B_{IJ} )
\sp I,J=1,2
\ee
The correlators are evaluated in the partition function of the
toroidally
compactified superstring
\be
\label{61}
Z =  \frac{1}{4}
\sum_{\bar{a}, \bar{b} =0}^1
(-)^{\ba + \bb + \m \ba \bb }
\th \bss^4
 (-)^{a+b+ab}
\th \ss^4
{\Ga_{2,2}(T,U) \over \t_2^3 |\eta|^{24}}
\ee
where $\m$ distinguishes between type IIA ($\m=0$) and type IIB
($\m=1$)
string theories. $\Ga_{2,2}$ is the usual $(2,2)$ lattice sum
describing
the possible wrappings of the string world-sheet on the target space
torus. Finally the correlation function has to be integrated over
the positions $z_i$ of the vertices on the world-sheet, and
over the world-sheet Teichm{\"u}ller parameter $\tau$ ( we
use the same notation as for the scalar modulus $\tau$,
but the context should make clear which one is meant), on the usual
fundamental domain ${\cal F}$ of the $SL(2,\Zint)$ modular group.
The different spin structures labelled by $a,b,\ba,\bb$ make distinct
contributions to the amplitude and have to be treated separately
according to  their parity $(-)^{ab}$.

$\bullet$ For even left and right moving spin structures, it can be
checked
from Riemann summation identity that terms with less that four
fermionic
contractions on both sides vanish after summing over even spin
structures.
This yields four powers of momenta on each side, and since we are
interested
in the leading ${\cal O}(p^8)$ contribution we can set $e^{ip\cdot
x}=1$.
The corresponding eight fermions on each side
have to be provided by the four gravitons vertex operators \ref{vop},
while only the bosonic part of \ref{mvop} can contribute.

The four pairs of fermions can be contracted in 60 different ways,
each of these giving, up to $g^{\m\n}$ factors, a contribution
\be
{1 \over 2} \sum_{a,b \; \rm even} (-)^{a+b} \th^4\ss
\prod_{i}^4
{ \th \ss(Z_i) \th_1'(0) \over \th \ss(0) \th_1(Z_i) }
= - {1\over 2} (2\p \eta^3) ^4
\label{rie}
\ee
where the result, after spin structure summation, no longer depends on
the contractions $Z_i$. On the other hand, the $\f$ modulus
insertion
yields a derivative with respect to $\f$:
\be
\label{8966}
\langle V_\phi(\bz_5,z_5) \rangle = {1\over \p\t_2} \pa_\f \Ga_{2,2}
\ee
again independent of $z^5$. Integrating over the positions of the vertices
according to $\int {\rd^2 z \over \t_2  }=1$, we obtain
\be
{\cal A}_{\bar{e}-e} = \T_{\bar{e},e}
\int {\rd^2 \t \over \t_2^2} {1\over 4}{1\over \t_2^3}
\left({\t_2 \over  \p}\right)^5
{ (2\p \eta^3)^4  (2\p \bar{\eta}^3)^4
\over |\eta|^{24} }
 {1\over \p\t_2} \pa_\f \Ga_{2,2}
\ee
where the kinematical factor $\T_{\bar{e},e}$ is provided by the
fermionic
contractions on each side:
\be
\T_{\bar{e},e}= 4~ t_8^{\bal_1 \bbe_1 \dots \bal_4 \bbe_4}
                   t_8^{\a_1 \b_1 \dots \a_4 \b_4}
\e^1_{\bal_1\a_1} p^1_{\bbe_1} p^1_{\b_1}
\dots
\e^4_{\bal_4\b_4} p^4_{\bbe_4} p^4_{\b_4}
\label{kinee}
\ee
We now use the momentum space representation of the Riemann tensor for
a
gravitational fluctuation around a flat background
\be
R_{\bal\bbe\a\b} =  {1\over 2} \left[ \e_{\bal\a} p_{\bbe} p_{\b}
- (\a\leftrightarrow\b) - (\bal\leftrightarrow\bbe)
  + ((\a,\bal)\leftrightarrow (\b,\bbe)) \right]
\ee
together with the symmetries
$(\a_i\leftrightarrow\b_i)$ of the $t_8$ tensor,
to rewrite
\be
\T_{\bar{e},e}= 16~
 t_8^{\bal_1 \bbe_1 \dots \bal_4 \bbe_4}
 t_8^{\a_1 \b_1 \dots \a_4 \b_4}
R_{\bal_1 \bbe_1 \a_1 \b_1}
\dots
R_{\bal_4 \bbe_4 \a_4 \b_4}
:= 16~t_8 t_8 R^4
\ee

$\bullet$ When one left or right spin structure is odd, the computation
is significantly different, since one vertex, say the modulus one,
has to be converted to the $(-1)$ ghost picture and supplemented by an
insertion
of a supercurrent. Effectively, in the $\rm \overline{odd}-odd$ case,
\be
V_\f \rightarrow v_{IJ} \bps(\bz)^I \ps(z)^J
{\rm e}^{i p \cdot X (\bz,z) } \;
G_F(0) \bar{G}_F(0)
\ee
where $G_F =  \pa X^\m \ps_\m + G_{KL} \pa X^K  \ps^L$.
The ten fermionic zero modes on both sides then have to be
saturated by the eight
fermions in the graviton vertices together with the two from the
modulus vertex and the supercurrent, while the integral over the
fermionic non-zero-modes induces the replacement
$\th  \ar{1}{1}^4  \ra (2\p \eta^3)^4$
in the partition function (\ref{61}).
The modulus insertion can be converted into a derivative with respect
to
 $\f$,
thanks to the supersymmetric partner of identity (\ref{8966}):
\be
\langle
v_{IJ}
\bps^I \ps^J
 G_{KL} \bpa X^K  \bps ^L
 G_{MN} \pa  X^M  \ps ^N
\rangle
=
{ \s_\f \over \p \t_2 } \pa_\f \Ga_{2,2}
\label{vevf}\ee
where $\s_\f=1$ for the $T,\bT$ moduli and $\s_\f=-1$ for the
$U,\bU$ moduli.
The ($\overline{o},o$) contribution therefore reduces to the same
contribution as the ($\overline{e},e$) one,  but for a kinematical
structure
\be
\T_{\bar{o},o}=
\e_8^{\bal_1 \bbe_1 \dots \bal_4 \bbe_4}
\e_8^{\a_1 \b_1 \dots \a_4 \b_4}
\e^1_{\bal_1\a_1} p^1_{\bbe_1} p^1_{\b_1}
\dots
\e^4_{\bal_4\a_4} p^4_{\bbe_4} p^4_{\b_4}
=
4~\e_8 \e_8 R^4
\label{kinoo}
\ee
and a crucial sign $(-)^{\m+\s_{\f}}$ depending on both the modulus
and the superstring we are considering. The fact that
the eight-dimensional Euler density $\e_8 \e_8 R^4$ is a total
derivative
should cause no concern here,
since this property is lost when its coefficient
becomes moduli-dependent.
Note also that in contrast to the four-dimensional case \cite{sixguys},
$t_8 t_8 R^4$ is now distinct from $\e_8 \e_8 R^4$, so that there
cannot
be any interferences between $\bar{e}-e$ and $\bar{o}-o$ amplitudes.

$\bullet$ When the spin structure is odd on one side and even on the
other
side, the modulus vertex has to be chosen in the (-1,0) picture. The
considerations of the previous cases apply on each side, and one
can again convert the modulus insertion into a derivative thanks to
\be
\eqalign{
\langle v_{IJ} \bps^I \pa X^J G_{KL} \bpa X^K \bps^L \rangle
=& ~i \s_\f \chi_\f \pa_\f \Ga_{2,2}
\cr
\langle v_{IJ} \bpa X^I \ps^J G_{MN} \pa X^M \ps^N \rangle
=& ~i \chi_\f \pa_\f \Ga_{2,2}
}
\ee
where $\chi_\f$ distinguishes chiral moduli ($\chi_T=\chi_U=1$)
from  antichiral ones ($\chi_{\bT}=\chi_{\bU}=-1$).
The result is again the same as in the $\bar{e}-e$ case, but for a
kinematical coefficient
\be 8~
\e_8^{\bal_1 \bbe_1 \dots \bal_4 \bbe_4}
\e_8^{\a_1 \b_1 \dots \a_4 \b_4}
R_{\bal_1 \bbe_1 \a_1 \b_1}
\dots
R_{\bal_4 \bbe_4 \a_4 \b_4}
:= 8~t_8 \e_8 R^4
\label{kineo}
\ee
and a prefactor $i(-1)^{\m}$ (resp. $i(-1)^{\m+\s_\f}$)
for the $\bar{e}-o$ (resp. $\bar{o}-e$) cases.

Putting all contributions together, we can write the scattering
amplitude as
\be
{\cal A} = 48\p
\left( t_8 + {i \over 2} (-)^{\m+\s_\f+\chi_\f} \e_8 \right)
\left( t_8 + {i \over 2} (-)^{\chi_\f}          \e_8 \right)
\pa_\f \int {\rd^2 \t \over \t_2}\Ga_{2,2}(T,U)
\label{full1ll}
\ee
where we fitted the overall coefficient to obtain the correct
decompactification limit.
The remaining fundamental domain integral was evaluated long ago in
\cite{dkl}:
\be
\int {\rd^2 \t \over \t_2} \Ga_{2,2}(T,U) =
-\log(T_2 |\eta(T)|^4 U_2 |\eta(U)|^4)
\label{dkli}
\ee
up to an irrelevant moduli-independent infrared ambiguity.

It is straightforward to integrate Eq.\ (\ref{full1ll})
with respect to  the moduli
$\f$ to
obtain the one-loop corrections to the
CP-even $t_8 t_8 R^4$ and $\e_8\e_8 R^4$ couplings. In the type IIB
case,
we obtain:
\be
\eqalign{
{\cal S}^{CP even}_{1-loop} = -2\p &\int \rd^8 x \sqrt{-g_\s} \left(
\left[ \log(T_2 |\eta(T)|^4) + \log( U_2 |\eta(U)|^4) t_8 t_8 R^4
\right]
\right. \cr
-& \left.
{1\over 4}
\left[ \log(T_2 |\eta(T)|^4) - \log( U_2 |\eta(U)|^4) \e_8 \e_8 R^4
\right]
\right)
}
\ee
However, in the CP-odd case, only the harmonic part of
Eq.\ (\ref{dkli}) can be integrated in the form of a moduli-dependent
$t_8 \e_8 R^4$ coupling, while the non-harmonic part $\log T_2 U_2$,
imputable to the IR divergence, has to be treated separately.
Let $X_7$ be the Chern-Simons form associated to the closed eight
form $t_8 \e_8 R^4 = 24 \Tr R^4 -6 (\Tr R^2)^2$: we can then rewrite
the CP-odd coupling of four gravitons and $T^2$ moduli as
(in the type IIB case)
\be
{\cal S}^{CP odd}_{1-loop} = 4\p
\int \rd^8 x \sqrt{-g_\s} \Im \left[\log\eta^4 (U)\right]~\e_8 t_8 R^4
-4\p \int \rd^8 x {1\over U_2} X_7 \wedge \rd U_1
\ee
The type IIA case is obtained under $T \leftrightarrow U$ exchange.

Going to the Einstein frame only modifies this result by higher
derivative
coupling to the dilaton.
We conclude that the tree-level
and one-loop corrections to
the $R^4$ thresholds for the eight-dimensional type IIB
string can be written as:
\be
\Delta_{tt}=2\zeta(3)V\tau_2^2-2\pi\log(T_2|\eta(T)|^4)
-2\pi\log(U_2|\eta(U)|^4)
\ee
\be
\Delta_{\e\e}=2\zeta(3)V\tau_2^2-2\pi\log(T_2|\eta(T)|^4)
+2\pi\log(U_2|\eta(U)|^4)
\ee
\be
\Theta=4\pi\Im[\log\eta(U)^4]
\ee

In particular,
the contribution of the world-sheet instantons of the fundamendal IIB
string to $\Delta_{tt}$
is given by
\be
{\cal I}_{1,0}=-8\pi {\rm
Re}~\log\left[\prod_{n=1}^{\infty}\left(1-e^{2\pi
inT}\right)\right]
\label{b4}\ee

\boldmath
\subsection{\bf Non-perturbative $(p,q)$-string instanton contribution}
\unboldmath

In addition to the fundamental string, the ten-dimensional type IIB
superstring theory possesses solitonic objects of various dimensions.
Wrapped on the compactification manifold, these configurations yield
instantons in lower dimensions. These instantons preserve one half
of the ten-dimensional supersymmetry, and therefore have the
correct number of fermionic zero-modes to contribute to $R^4$ couplings.
The D-instanton is localized to a point in spacetime and yields
the same contribution in any compactification of type IIB theory,
up to a volume factor. The D3,5,7-branes and the NS 5-brane
contribute for $D\le 6,4,2,4$ respectively.
On the other hand,
the D-strings start contributing for $D=8$, where they can
supersymmetrically wrap around the two-torus.
The D-strings have charges $(p,q)$ under
the NS-NS and R-R antisymmetric tensors
($p,q$ are coprime: $(p,q)=1$), and form a multiplet
under $SL(2,\Zint)_{\tau}$ symmetry. The $(1,0)$ string corresponds
to the fundamental type IIB string, and its contribution is
known from the perturbative one-loop computation (\ref{b4}).
One can then apply $SL(2,\Zint)_{\tau}$ to infer the contributions
of all $(p,q)$ strings.

The world-sheet Nambu-Goto action of a (0,1) D-string is
known to be
\cite{pol}:
\be
S_{0,1}={e^{-\phi}\over 2\pi}\int d^2\s \sqrt{{\rm det}(\hat G+{\cal
F})}+{i\over
2\pi}\int\hat B_R
\label{bb5}\ee
for vanishing background Ramond scalar expectation value.
The hat denotes pulled-back quantities: $\hat
G_{\a\b}=G_{\mu\n}\pa_{\a}
X^{\mu}\pa_{\b}X^{\n}$, etc.
and ${\cal F}=F-\hat B_N$, with $F_{\a\b}$ the field strength of the
world-sheet gauge field.
When the background scalar $a=\tau_1$ is switched on, this becomes
\be
S_{0,1}={|\tau|\over 2\pi}\int d^2\s \sqrt{{\rm det}(\hat G+{\cal
F})}+{i\over2\pi}\int\left( \hat B_R + \tau_1 {\cal F} \right)
\label{bb6}\ee
where the $\tau_1 {\cal F}$ coupling ensures
anomaly cancellation \cite{ghm}.

Using Cartesian coordinates $X^1,X^2\in [0,2\pi]$ for the target space
torus and
$\sigma_{1,2}\in [0,2\pi]$ for the D1-brane,
the $\sigma$-frame target metric is given in Eq.\ (\ref{7}).
The supersymmetric embedding wrapping the string world-sheet around
the two-torus can be written as
\be
\left(\matrix{X^1\cr X^2\cr}\right)=\left(\matrix{m_1&n_1\cr
m_2&n_2\cr}\right)\left(\matrix{\sigma_1\cr \sigma_2\cr}\right)
\label{b6}\ee
A non-degenerate orientation preserving mapping is obtained
$N=m_1n_2-m_2n_1>0$, while
for $m_1n_2-m_2n_1 <0$, the orientation is reversed and the induced
complex structure is complex-conjugated.
The equations of motion also require ${\cal F}=$constant.
Setting the constant to zero correcponds to the (0,1) string \cite{pq}. 
Evaluating the action on this instanton configuration, we obtain
\be
S_{0,1}^{\rm class}=2\pi|N||\tau|T_2+2\pi
i NB_R
\label{b7}
\ee
This implies that the $(0,1)$-string instanton has an effective
$T$-modulus given by
\be
T_{0,1}=B_R+i|\tau|T_2
\ee
Using the SL(2,$\Z)$ symmetry we obtain that the effective modulus
of a $(p,q)$ string is
\be
T_{p,q}=(qB_R-pB_N)+i|p+q\tau|T_2
\label{b8}\ee
Thus, the contribution from all $(p,q)$ D-strings can be written as
\be
{\cal I}_{p,q}=-8\pi {\rm
Re}~\log\left[\prod_{n=1}^{\infty}\left(1-e^{2\pi
inT_{p,q}}\right)\right]
\label{b9}
\ee
Together with the D-instanton contribution (\ref{a6}), we obtain our
conjecture for the exact $R^4$ thresholds in 8 dimensions:
\be
\label{b10}
\Delta_{tt}=V\sqrt{\tau_2}f_{10}(\tau,\bar\tau)-2\pi\log
T_2-2\pi\log(U_2|\eta(U)|^4)+\sum_{(p,q)=1}{\cal I}_{p,q}
\ee
\be
\label{b11}
\Delta_{\e\e}=V\sqrt{\tau_2}f_{10}(\tau,\bar\tau)-2\pi\log
T_2+2\pi\log(U_2|\eta(U)|^4)+\sum_{(p,q)=1}{\cal I}_{p,q}
\ee
while we do not expect any corrections to the already duality-invariant
$t_8 \e_8 R^4$ coupling. In order to be acceptable, this
result should satisfy the requirement of $SL(3,\Zint)\ti SL(2,\Zint)_U$
invariance. The invariance under $SL(2,\Zint)_U$
is already incorporated in the above Equations.
Indeed, $\log(U_2|\eta(U)|^4$ is the order-1 Eisenstein
series for $SL(2,\Zint)_{U}$, and the other terms are invariant
under that group.
We will show in the next Section the remaining terms
can be rewritten in terms of the
$SL(3,\Zint)$ Eisenstein order-3/2 series, which will
make the invariance under $SL(3,\Zint)$ obvious. This series is actually
logarithmically divergent, and it will turn out
necessary to add to Eqs.\ (\ref{b10},\ref{b11}) an extra
logarithmic contribution not captured
by perturbation theory.

\boldmath
\section{$SL(3,\Zint)$ invariance of the $R^4$ thresholds}
\setcounter{equation}{0}
\unboldmath
In view of the ten-dimensional result (\ref{a5}), it is natural
to conjecture that the eight-dimensional $t_8 t_8 R^4$
threshold can be written in terms
of the order-$s=3/2$ Eisenstein  series for $SL(3,\Zint)$.
This will fulfill the requirements of $SL(3,\Zint)$ invariance
and ten-dimensional decompactification limit. In fact we shall show
here that this series gives precisely the result motivated in the
previous Section.

The Eisenstein $SL(3,\Zint)$ series with order-$s$ is defined as
\be
E_{s}\equiv \hat{\sum_{m_i\in
Z}}\left(\sum_{i,j=1}^3m_iM^{ij}m_j\right)^{-s}
= \hat{\sum_{m_i\in \Zints}}\n^{-s/3}\left[{|m_1+m_2\tau+m_3B|^2\over
\tau_2}+{m_3^2\over \n}\right]^{-s}
\label{17}\ee
where $\hat{\sum}$ stands for the sum with (0,0,0) omitted.
$E_{s}$ is by construction $SL(3,\Zint)$-invariant.

Introducing the Laplacian operator on the $SL(3,\Real)/SO(3)$
homogeneous space
\be
\Delta=4\tau_2^2\pa_{\tau}\pa_{\bar\tau}+{1\over
\n\tau_2}\left|\pa_{B_N}-\tau\pa_{B_R}\right|^2+3\pa_{\n}(\n^2\pa_{\n})
\label{18}\ee
we deduce that $E_s$ is an eigenfunction of the Laplacian
\be
\Delta E_s={2s(2s-3)\over 3} E_s
\label{19}\ee
For $s>3/2$, $E_s$ is an absolutely convergent series. $E_{3\over 2}$ is
logarithmically divergent and is also annihilated by the Laplacian.
This is the relevant function for the threshold since we
already know
that there is a physical logarithmic divergence in the
eight-dimensional
$t_8t_8R^4$ term at one-loop.
It also matches the ten-dimensional conjecture \cite{vg1}, as well
as the recent M-theory motivated proposal \cite{vg2}.
In  the sequel we will use $\zeta$-function regularization by keeping
$s$ arbitrary and larger then 3/2 where the sum converges.
We use an ``$\overline{MS}$"-like definition:
\be
\hat E_{3\over 2}=\lim_{\epsilon\to 0^+}\left[E_{{3\over
2}+\epsilon}-{2\pi\over \epsilon}-4\pi(\gamma-1)\right]
\label{199}\ee
where $\gamma$ is the Euler constant.
Because of this subtraction, $\hat E_{3\over 2}$ is no longer  a
zero-mode
of the Laplacian, but instead satisfies
\be
\Delta\hat E_{3\over 2}=4\pi
\label{1999}\ee

We now show that this function contains all the contributions
expected
from our previous arguments.
We introduce the integral representation
\be
E_s(M)={\pi^s\over \Gamma(s)}\int_0^{\infty}{dt\over t^{1+s}}
{}~\hat{\sum_{m_i\in \Zints}}\exp\left[-{\pi\over t}
\left(\sum_{i,j=1}^3m_iM^{ij}m_j\right)\right]
\label{200}\ee
$$=\n^{-s/3}{\pi^s\over \Gamma(s)}\int_0^{\infty}{dt\over t^{1+s}}
{}~\hat{\sum_{m_i\in \Zints}}\exp\left[-{\pi\over t} \left({m_3^2\over
\n}+{|m_1+m_2\tau+m_3B|^2\over \tau_2}\right)\right]
$$
In Appendix A, we evaluate this integral for arbitrary s. Here we
will set $s=3/2$ and mention at the appropriate point the modification
from the regularization.

We first split the sum as $\hat{\sum}_{m_i\in
Z}=\sum_{m_1\not=0,m_2,m_3=0}+
\sum_{m_1\in \Zints}\hat{\sum}_{m_{2,3}\in \Zints}$ to obtain
\be
\hat E_{3\over2}=4\pi\n^{-1/2}\sum_{m_1=1}^{\infty}\int_{0}^{\infty}
{dt\over t^{5/2}}\exp\left[-{\pi m_1^2\over
t\tau_2}\right]+J_1=I_0+J_1
\label{21}\ee
where
\be
I_0=2{\tau_2^{3/2}\over \n^{1/2}}\z(3)
\label{23}
\ee
\be
J_1=2\pi\n^{-1/2}\int_{0}^{\infty}
{dt\over t^{5/2}}\hat{\sum_{m_{2,3}\in \Zints}}\sum_{m_1\in \Zints}
\exp\left[-{\pi\over t} \left({m_3^2\over
\n}+{|m_1+m_2\tau+m_3B|^2\over \tau_2}\right)\right]
\label{22}
\ee
We now Poisson-resum on $m_1$, change variable $t\to 1/t$ and then
separate the $m_1=0$ and $m_1\not=0$ contributions to obtain
$J_1=J_2+J_3$, with
$$
J_3=2\pi{\sqrt{\tau_2}\over \sqrt{\n}}\int_0^{\infty}dt
{}~\sum_{m_1\not=0}\hat{\sum_{m_{2,3}\in \Zints}}\exp\left[
-{\pi\tau_2m_1^2\over t}-{\pi t\over \tau_2}(m_2\tau_2+m_3B_2)^2-
\right.$$
\be
\left.-{\pi t m_3^2\over \n}-2\pi im_1(m_2\tau_1+m_3B_1)\right]
\label{24}\ee
where $B=B_R+\tau B_N=B_1+iB_2$
and
\be
J_2=2\pi{\sqrt{\tau_2}\over
\sqrt{\n}}\int_0^{\infty}dt~\hat{\sum_{m,n\in
Z}}\exp\left[-\pi t\tau_2|m+nT|^2\right]
\label{25}\ee
In the above Equation, we changed variables to the T modulus using
Eq.\ (\ref{888},\ref{10}).$J_2$ is in fact the piece
responsible for the IR divergence.
In the Appendix we show that the regulated expression is
\be
J_{2,\rm reg}=-\pi\left(\log\tau_2+{1\over
3}\log\n\right)-2\pi\log T_2+{2\pi^2\over 3}T_2+I_{1,0}
\label{31}\ee
$$=-\pi\left(\log\tau_2+{1\over
3}\log\n\right)-2\pi\log\left[T_2|\eta(T)|^4\right]
$$
where
\be
I_{1,0}=8\pi {\rm Re}~\sum_{m,n=1}^{\infty}{1\over n}e^{2\pi i
mnT}=-8\pi
{\rm Re}~\log\left[\prod_{m=1}^{\infty}\left(1-e^{2\pi
imT}\right)\right]
\label{29}\ee
is precisely the contribution of the fundamental string
world-sheet instantons.

We will now proceed to evaluate the left-over integral $J_3$.
We will again split the summation as
$(m_2,m_3)'=(m_2\not=0,m_3=0)+(m_2,m_3\not=0)$ and Poisson-resum over
$m_2$
in the second sum to obtain $J_3=I_D+J_4$ with
\be
I_D=8\pi\sqrt{\tau_2\over \n}\sum_{p\not=0}\sum_{n=1}^{\infty}
\left|{p\over n}\right|K_{1}(2\pi\tau_2 |p|n)e^{2\pi ipn\tau_1}
\label{32}\ee
where $K_1$ is the standard Bessel function, and
\be
J_4=2\pi\sum_{m_{1,3}\not=0}\sum_{m_2\in \Zints}{1\over |m_3|}\exp\left[
-2\pi|m_3||m_2+m_1\tau|T_2-2\pi i{m_3\over \tau_2}\left(
m_2B_2+m_1(\tau_1B_2-B_1\tau_2)\right)\right]
\label{33}\ee
The piece $I_D$ is the contribution of the D-instantons
conjectured in Ref. \cite{vg1}.

We now split the summation over $m_2$ into the $m_2=0$ and $m_2\not =0$
pieces to write $J_4=I_{0,1}+J_5$
with
\be
{\cal I}_{0,1}=-8\pi {\rm Re}~\log\left[\prod_{m=1}^{\infty}\left(1-e^{2\pi
im\left(
B_R+i|\tau|T_2\right)}\right)\right]
\label{34}\ee
We recognize in ${\cal I}_{0,1}$ the
contribution of the (0,1) D1-string instantons. Pursuing further,
\be
J_5=2\pi\sum_{m_i\not=0}{1\over |m_3|}\exp\left[
-2\pi|m_3||m_2+m_1\tau|T_2+2\pi im_3\left(m_1B_R-m_2B_N
\right)\right]
\label{35}\ee
$$
=-4\pi Re \sum_{m_{1,2}\not=0}\log\left(1-\exp\left[
-2\pi|m_2+m_1\tau|T_2+2\pi i\left(m_1B_R-m_2B_N
\right)\right]\right)
$$
where we have used the definition $B=B_R+\tau B_N$.
If we denote by $n$ the (positive) greatest common divisor from
any pair of non-zero integers $(m_1,m_2)$, then we can write
$\{m_1,m_2\}=n\{p,q\}$
with $(p,q)=1$.
Moreover  $(p,q)$ and $-(p,q)$ give the same contribution.
Summing over $(p,q)$ modulo this charge conjugation, we
finally obtain
\be
J_5=\sum_{p,q\not=0,(p, q)=1}~ {\cal I}_{p,q}
\label{36}\ee
with
\be
{\cal I}_{p,q}=-8\pi {\rm  
Re}\log\left[\prod_{n=1}^{\infty}\left(1-e^{2\pi in
T_{p,q}}\right)\right]
\label{37}\ee
and
\be
T_{p,q}=(qB_R-pB_N)+i|p+q\tau|T_2
\label{38}\ee
Those are recognized as the contributions of
the dyonic $(p,q)$-string instantons (\ref{b9}).

Putting everything together we obtain the following expansion for
the $SL(3,\Zint)$ Eisenstein order-3/2 series:
\be
\hat E_{3\over 2}=2{\tau_2^{3/2}\over \n^{1/2}}\z(3)+{2\pi^2\over 3}T_2
+4\pi\log\n^{1/3}+I_D+\sum_{(p,q)=1}{\cal I}_{p,q}
\label{59}\ee
This reproduces the results announced in Eqs\ (\ref{b10},\ref{b11}) ,
up to the logarithmic term $4\pi\log\n^{1/3}$ which is
required for $SL(3,\Zint)$ invariance.
We therefore conclude that the exact eight-dimensional thresholds
can be written as:
\be
\Delta_{tt}=\hat E_{3\over 2}-2\pi\log(U_2|\eta(U)|^4)\;\;\;,\;\;\;
\Delta_{\e\e}=\hat E_{3\over 2}+2\pi\log(U_2|\eta(U)|^4)\label{}\ee

\boldmath
\section{Compactification of type IIB String Theory to $D<8$}
\unboldmath

In this Section we shall investigate how the non-perturbative result
of $D=8$ can be extended to lower dimensions, focusing mainly on the
seven-dimensional case. We will again propose
a $U$-duality invariant expression for $R^4$ thresholds.
We will show that it reproduces tree-level, D-instanton, fundamental
string and
$(p,q)$-string instanton contributions.
The six-dimensional case will also be briefly discussed,
where additional three-brane contributions are expected.

\subsection{\bf Perturbative compactification on a $N$-torus}
Let $G_{ij}$ be the $\sigma$-frame metric of the $N$-torus, and
$B^\a_{ij}$ the associated antisymmetric tensors ($i=1$ stands for the
R-R antisymmetric tensor, $i=2$ for the NS-NS one).
We will separate the overall volume as
\be
G_{ij}=V^{2/N}\tilde G_{ij}\;\;\;,\;\;\;\sqrt{{\rm
det}G}=V\;\;\;,\;\;\;
{\rm det}\tilde G=1
\label{c1}\ee
and define the $SL(2,\Zint)_\tau$ invariant scalar
\be
\n={1\over \tau_2V^{4/N}}
\label{c2}\ee
The effective action of the IIB superstring toroidally compactified
to $10-N$ dimensions is, in the Einstein frame,
\be
S_{10-N}={1\over 2k_{10-N}^2}
\int d^{10-N}x\sqrt{-g_E}\left[R-{N\over 2(8-N)}\left({\pa
\n\over \n}\right)^2-{1\over 2}{\pa\tau\pa\bar\tau\over
\tau_2^2}+{1\over 4}Tr(\pa\tilde G\pa\tilde G^{-1})-\right.
\label{c3}\ee
$$\left.-{\n\over 4}\tilde G^{ik}
\tilde G^{jl}{(\pa B^1_{ij}+\tau \pa B^2_{ij})
(\pa B^1_{kl}+\bar\tau \pa B^2_{kl})\over \tau_2}
-{\nu^2\over 2.4!}\tilde G^{im}\tilde G^{jn}\tilde G^{kp}\tilde G^{lq}
\pa C_{ijkl} \pa C_{mnpq}
+\cdots\right]
$$
with $\k_{10-N}=\k_{10}/(2\pi)^{N/2}$.
Here, we reinstated the four-form, which gives rise to moduli for $N\ge 4$.

We parametrize the $t_8 t_8 R^4$ threshold as:
\be
S^{R^4}={\cal N}_{10-N}\int d^{10-N} x\sqrt{-g_E}\left[
\Delta_{tt}t_8t_8R^4\right]
={\cal N}_{10-N}\int d^{10-N}x\sqrt{-g_\sigma}~
\nu^{2N-4 \over 8-N} V^{1-{2\over N}}
\left[\Delta_{tt}t_8t_8R^4\right]
\label{b100}
\ee
where ${\cal N}_{10-N}=(2\pi)^N{\cal N}_{10}$.
The one-loop perturbative correction can again be written in terms
of the $(N,N)$ torus lattice sum:
\be
\nu^{2N-4 \over 8-N} V^{1-{2\over N}}
\Delta^{1-loop}_{tt}= I_{N,N}
=2\pi\int_{\cal F}{d^2\tau\over
\tau_2^2}\tau_2^{N\over 2}\Gamma_{N,N}
(G,B_N)
\label{60}\ee
where
\be
\tau_2^{N\over 2}\Gamma_{N,N}=\sqrt{G}\sum_{m^i,n^i\in \Zints}\exp\left[
-{\pi\over
\tau_2}(G_{ij}+B_{N;ij})(m^i+n^i\tau)(m^j+n^j\bar\tau)\right]
\label{62}\ee
This integral has an infrared power divergence. It can
be evaluated by the method of orbits
\footnote{We would like to thank M. Henningson
for collaborating in the calculation of this integral.} \cite{dkl}.
Define the following sub-determinants, $d^{ij}=m^i n^j-m^j n^i$.
$d^{ij}$ is an $N\times N$ antisymmetric matrix.
The action of the Teichm{\"u}ller group $SL(2,\Zint)$
decomposes into the following orbits:

$\bullet$ The trivial orbit, $m^i,n^i=0$, with a contribution
\be
I^{tr}_{N,N}=2\pi \sqrt{G} \int_{\F} {d^2 \tau \over \tau_2^2} 1
={2\pi^2\over 3}V
\label{646}\ee

$\bullet$ The degenerate orbits, with all $d$'s being zero.
In this case we can set $n^i=0$ and unfold the integration domain
$\F$ onto the strip $\t_1\in[-{1\over 2},{1\over 2}],\t_2\in \Real^+$.
We obtain
\be
I^{d}_{N,N}=2V \hat{\sum_{m^i}}{1\over m^{i}
G_{ij}m^{j}}=2~V^{1-{2\over N}}\hat{\sum_{m^i}}{1\over m^{i}
\tilde G_{ij}m^{j}}
\label{65}\ee
Note that the sum is, up to a volume factor,
the Eisenstein series $E_1(\tilde G)$
for $SL(N,\Zint)$.
It is indeed power-divergent for $N>2$ and in the Appendix we show
(for the $N=3$ case) how the divergence can be subtracted. In the
sequel we assume that this is carried out.

$\bullet$ The non-degenerate orbits, where at least one of the $d^{ij}$
is non-zero. We can completely fix the modular $SL(2,\Zint)$ action
\be
\left(\matrix{m^i\cr n^i\cr}\right)\to \left(\matrix{a& b\cr c&
d\cr}\right)
 \left(\matrix{m^i\cr n^i\cr}\right)
\label{67}\ee
in order to unfold the integration domain to twice the upper-half
plane.
After Gaussian integration on $\tau$, we obtain:
\be
I^{n.d}_{N,N}=4\pi V^{1-{2\over N}}\sum {\exp\left[-2\pi
V^{2/N}\sqrt{(m\cdot\tilde
G\cdot m)(n\cdot\tilde G\cdot n)-(m\cdot\tilde G\cdot n)^2}+2\pi
i(m\cdot B_N\cdot n)\right]\over \sqrt{(m\cdot\tilde G\cdot
m)(n\cdot\tilde G\cdot n)-(m\cdot\tilde G\cdot n)^2}}
\label{66}\ee
The summation is done over all sets of 2N integers, having at least
one non-zero $d^{ij}$, modded out by the modular
$SL(2,\Zint)$ action.
This part is IR-finite.

\subsection{\bf Fundamental string world-sheet instantons on $T^3$}
For $N=3$, we can ``dualize" $B^\a_{ij}=\e_{ijk}B^{\a k}$, where
$\e_{ijk}$ is the antisymmetric Levi-Civita symbol with $\e_{123}=1$,
and write the action in the simpler form:
\be
S_{7}={1\over 2k_7^2}
\int d^{7}x\sqrt{-g_E}\left[R-{3\over 10}\left({\pa \n\over
\n}\right)^2-{1\over 2}{\pa\tau\pa\bar\tau\over \tau_2^2}+{1\over
4}{\rm Tr}(\pa\tilde G\pa\tilde G^{-1})-\right.
\label{c4}\ee
$$\left.-{\n\over 2}\tilde G_{ij}{(\pa_{\mu}B^{1i}+\pa_{\mu}\tau
B^{2i})(\pa_{\mu}B^{1j}+\bar\tau\pa_{\mu}
B^{2j})\over \tau_2}+\cdots\right]
$$
Again the scalar kinetic terms can be written in the form Tr$(\pa M\pa
M^{-1})/4$, where $M$ is a symmetric $5\times 5$ matrix with
determinant 1, parametrizing the $SL(5,\Real)/SO(5)$
coset:
\be
M=\n^{3/5}\left(\matrix{g_{\a\b}&g_{\a\b}B^{\b j}\cr
 g_{\a\b}B^{\b i}&{\tilde G^{ij}\over
\nu}+B^{\a i}B^{\b j}g_{\a\b}\cr}\right)
\;\;\;,\;\;\;g_{\a\b}={1\over \tau_2}\left(\matrix{1&\tau_1\cr
\tau_1&|\tau|^2\cr}\right)\label{c5}\ee
The $U$-duality group is
$SL(5,\Zint)\supset O(3,3,\Zint) \times SL(2,\Zint)_{\tau}$, and
acts as $M\to \Omega M\Omega^T$.

Using the dual representation of $B_N$ we can also rewrite the contribution
of the non-degenerate orbit to the one-loop threshold in a way
that will be crucial for comparison with the full
non-perturbative result. It amounts to trading the summation
over the $m,n$ integers modulo $SL(2,\Zint)$ for a summation over
the $SL(2,\Zint)$ invariant integers $d^{ij}$, which can similarly
be dualized as $d^{ij}=\e^{ijk}d_k$.
With these notations, one can show that
\be
(m\cdot\tilde
G\cdot m)(n\cdot\tilde G\cdot n)-(m\cdot\tilde G\cdot n)^2=d_i(\tilde
G^{-1})^{ij}d_j\;\;\;,\;\;\;m\cdot B^N \cdot n=d_i B_N^i
\label{70}\ee
We can then rewrite the contribution of the non-degenerate orbit as
\be
I^{n.d}_{3,3}=4\pi V^{1/3}\overline{\sum}_{d_i}{\exp\left[-2\pi
V^{2/3}\sqrt{d\cdot
\tilde G^{-1}
\cdot d}+2\pi id\cdot B^N\right]\over \sqrt{d\cdot \tilde G^{-1}
\cdot d}}
\label{71}\ee
This is the contribution of the fundamental $(1,0)$ string
world-sheet instantons.
There is a lot hidden in the summation sign, $\overline{\sum}_{d_i}$.
We will make it more explicit presently, distinguishing the
following three cases:

$\bullet$ Only one of the $d_i$ is non-zero, say $d_1\not=0$,
$d_2=d_3=0$.
Then we can fix the $SL(2,\Zint)$ action by choosing the following
representatives:
\be
\left(\matrix{m_1&m_2&m_3\cr n_1&n_2&n_3\cr
}\right)=\left(\matrix{0&k&j\cr 0&0&p\cr }\right)
\ee
with $d_1=kp$, $p>0$, $0\le j <p$.
The sum in Eq.\ (\ref{71}) becomes
\be
I^{n.d(1)}_{3,3}=4\pi V^{1/3}\sum_{k\not=0,p>0\atop j mod(p)}
{\exp\left[-2\pi V^{2/3}|kp|\sqrt{\tilde G^{11}}+2\pi i
kpB^1\right]\over
|kp|\sqrt{\tilde G^{11}}}
\label{81}\ee
$$
=-8\pi {V^{1/3}\over \sqrt{\tilde G^{11}}}
\sum_{p=1}^{\infty}\Re \log\left[1-\exp\left(-2\pi
V^{2/3}|k|\sqrt{\tilde G^{11}}+2\pi i kB_N^1\right)\right]
$$

$\bullet$ Two out of the three $d_i$ are non-zero, say $d_1,d_2$.
Then we can choose the following representative:
\be
\left(\matrix{m_1&m_2&m_3\cr n_1&n_2&n_3\cr
}\right)=\left(\matrix{k_2&k_1&j\cr 0&0&p\cr }\right)
\ee
with $d_1=k_1p$, $d_2=k_2p$, $p>0$, $0\le j < p$.
The sum in Eq.\ (\ref{71}) becomes in this case
\be
I^{n.d(2)}_{3,3}=4\pi V^{1/3}\sum_{k_i\not=0,p>0\atop j mod(p)}
{\exp\left[-2\pi V^{2/3}p\sqrt{k_i \tilde G^{ij} k_j}+2\pi i
pk_iB_N^i\right]\over p\sqrt{k_i \tilde G^{ij}k_j}}
\label{82}\ee
$$
=2\pi\sum_{k_i\not=0} {V^{1/3}\over \sqrt{k_i\tilde G^{ij}
k_j}}\sum_{p=1}^{\infty}\exp\left(-2\pi
V^{2/3}p\sqrt{k_i \tilde G^{ij}k_j}+2\pi i pk_iB_N^i\right)
$$

$\bullet$ Finally, consider the case where all the $d_i$ are non-zero.
Then we can choose the following representative
\be
\left(\matrix{m_1&m_2&m_3\cr n_1&n_2&n_3\cr
}\right)=\left(\matrix{m_1&m_2&m_3\cr 0&n_2&n_3\cr }\right)
\ee
with $n_2>0$, $0\le m_3 < n_3$.
$d_1=m_2n_3-m_3n_2$, $d_2=m_1n_3$, $d_3=m_1n_2$.
In this case we can show that
 a given $(d_1,d_2,d_3)$ with greatest
common divisor N corresponds to $\sum_{p|N}~p$ equivalence
classes of integers $n,m$.
Thus we can write the contribution as
\be
I^{n.d(3)}_{3,3}=2\pi V^{1/3}\sum_{N=1}^{\infty}\sum_{p|N}{p\over
N}\sum_{(d_{1}, d_{2},d_{3})=1} {\exp\left[-2\pi V^{2/3}N\sqrt{d\cdot
\tilde G^{-1}\cdot d}+2\pi i Nd\cdot B\right]\over \sqrt{d\cdot
\tilde G^{-1}\cdot
d}}
\label{83}\ee
$$
=-2\pi\sum_{(d_{1}, d_{2}, d_{3})=1} {V^{1/3}\over \sqrt{d\cdot
\tilde G^{-1}\cdot d}}\sum_{p=1}^{\infty}\log\left[1-\exp\left(-2\pi
V^{2/3}p\sqrt{d\cdot \tilde G^{-1}\cdot d}+2\pi i pd\cdot
B\right)\right]
$$

\subsection{\bf Exact gravitational thresholds in $D=7$ and $D=6$}
In addition to the perturbative contributions of the fundamental
type IIB string, we expect instanton contributions from D-instantons
and $(p,q)$-strings wrapped on the 3-torus.
As in eight dimensions we can write down the result since the
D-instanton contribution is known from
the ten-dimensional result, and the $(p,q)$
instanton contributions
can be obtained from the contribution of the fundamental $(1,0)$
string in Eq.\ (\ref{71}).
Thus we expect that
\be
\Delta^7_{tt}=\n^{-9/10}f_{10}(\tau,\bar\tau)+2 \nu^{-2/5}
E^{SL(3)}_{1,\rm reg}(\tilde G)
+\sum_{(p,q)=1}{\cal I}^7_{p,q}
\label{73}\ee
where
\be
{\cal I}^7_{p,q}=2\pi\n^{-2/5}\overline{\sum}_{d_i}{\exp\left[
-2\pi|p+q\tau|{\sqrt{d\cdot\tilde G^{-1}\cdot d}\over
\sqrt{\tau_2\nu}}+2\pi i
d\cdot\left(q B_R-p B_N\right)\right]\over \sqrt{d\cdot
\tilde G^{-1}\cdot d}}
\label{74}\ee

We now show that, unsurprisingly, the full non-perturbative
threshold (\ref{73}) can be written in terms of the order-3/2
Eisenstein series for $SL(5,\Zint)$. This series is
defined in terms of the $SL(5,\Real)/SO(5)$ matrix $M$ in Eq.\ (\ref{c5}) 
as:
\be
E_{3\over 2}^{SL(5)}(M)=2\pi\int_{0}^{\infty}
\left(  {1\over
t^{5/2}} \hat{\sum_{m^i}} 
\exp\left[-{\pi\over
t}m^iM_{ij}m^j\right]- 1
\right)\, dt
\ee
$$
=2\pi\int_{0}^{\infty} \left( {1\over t^{5/2}}
\hat{\sum_{m^i,n_{\a}}} \exp\left[-{\pi\n^{3/5}\over t}\left({|m_1+\tau m_2 +
(B_R+\tau B_N)\cdot n |^2 \over \tau_2}
+{n\cdot\tilde G^{-1}\cdot n\over \n}
\right)\right]
-1\right) dt
$$
Using the integral representation, and going through the same steps
as in the $SL(3)$ case, we can establish that it is equal (up to an
additive constant) to
\be
E_{3\over 2}^{SL(5)}=\n^{-9/10}f_{10}(\tau,\bar\tau)+2\nu^{-2/5}
E^{SL(3)}_{1,\rm reg}(\tilde G)
+\sum_{(p,q)=1}\hat{\cal I}^7_{p,q}
\label{86}\ee
where
\be
\hat{\cal I}^7_{p,q}=2\pi\nu^{-2/5}\sum_{l=1}^{\infty}\hat{\sum}_{n^i}
{\exp\left[
-2\pi l |p+q\tau|{\sqrt{n\cdot\tilde G^{-1}\cdot n}\over
\sqrt{\tau_2\nu}}+2\pi i
l n\cdot\left(q B_R-q B_N\right)\right]\over \sqrt{n\cdot
\tilde G^{-1}\cdot n}}
\label{87}\ee
Separating the three cases, corresponding to one, two or three non-zero
$n^i$, and taking out the greatest common divisor in the last case, we
observe
that
\be
\hat{\cal I}^7_{p,q}={\cal I}^7_{p,q}
\ee
and Eq.\ (\ref{86}) coincides with Eq.\ (\ref{73}), which proves our
claim.
This concludes our discussion of the seven-dimensio\-nal case.

We now briefly turn to the six-dimensional case. There we expect
D3-instanton corrections in addition to the ones existing in higher
dimensions.
The scalar manifold is now $SO(5,5,\Real)/(SO(5)\times SO(5))$
and the $U$-duality group $SO(5,5,\Zint)$. It is more convenient
to parametrize this manifold in terms of type IIA variables.
Indeed, type IIA string compactified on $T^4$ can be viewed
as the eleven-dimensional M-theory compactified on a
five-torus.
The scalars are the five-dimensional metric $G_{ij}$ and the internal
components
of the three-form $C_{ijk}$, which can be dualized into a two
index antisymmetric tensor
$C_{ijk}= \e_{ijklm} \tilde C^{lm}$.
We can now construct the standard $10\times 10$ symmetric $SO(5,5)$  matrix
\be
M_{6}=\left(\matrix{\det G \, G^{-1} - \tilde C G \tilde C &\tilde G \tilde C\cr 
-\tilde C G& G\cr}\right)\ ,
\label{79}\ee
in units of the 11D Planck scale.
We again conjecture that the threshold should be given by the
order 3/2 Eisenstein series, appropriately
regularized:
\be
{\Delta^6_{tt}\over 2\pi}=\int_{0}^{\infty}
\left( {1\over
t^{5/2}} \hat{\sum_{m^i,n^i}}
\exp\left[-{\pi\over t}\left( (m^i + \tilde C^{ik} n_k) G_{ij} (m^j + \tilde C^{jl} n_l) + (\det G) \, n_i G^{ij} n_j \right)\right]- t^{5/2}\right) dt
\label{80}\ee
It remains to be shown that this expression reproduces
the contribution of the ten-dimension\-al instantons, the
D$(p,q)$-strings (that we can obtain from a one-loop
$(1,0)$ string calculation), plus an extra piece that will
be attributed to D3-brane instantons.
We leave the further analysis of the six-dimensional case to a future
publication.

\section{Conclusions}

In this paper we analysed in detail the threshold corrections
of various $R^4$ terms in type IIB string theory compactified to eight
and seven dimensions.
The $R^4$ terms are BPS-saturated and receive perturbative
contributions from one loop only, and from short N=8 multiplets.
In ten dimensions, in addition to the perturbative contributions there are
D-instanton corrections.
In eight dimensions, $(p,q)$-string instantons can also contribute.
We have calculated their contribution and shown that the full result is
a
order-3/2 Eisenstein
series for $SL(3,\Zint)$, plus a order-one
Eisenstein series for $SL(2,\Zint)$,
invariant under the
$SL(3,\Zint)\times SL(2,\Zint)$ $U$-duality group in eight dimensions.
This is in agreement with the M-theory calculation of \cite{vg2}.
Furthermore, it was noticed that
$SL(3,\Zint)$ invariance requires the presence of a logarithmic term, absent
in perturbation theory or in the instanton calculations.
Its presence is
tied to the logarithmic infrared divergence of the threshold in eight
dimensions.

In seven dimensions the $U$-duality group is $SL(5,\Zint)$. The same type of
instantons contribute as in the eight-dimensional case.
We have calculated the exact threshold in this case
and shown that it is given by a order-3/2 series for $SL(5,\Zint)$.

A more interesting case is the six-dimensional one, where D3-instantons
are expected to contribute.
We do expect again that the threshold will be given by a order-3/2
form ofor $SO(5,5,\Zint)$.
This conjecture is also valid in lower dimensions, where one has to
consider the order-3/2 Eisenstein  series
for the exceptional
$E_{(n,n)}(\Zint)$ discrete groups.
Checking this conjecture for $D\le 6$ promises interesting
insight into instanton calculus in string theory.

We make a final comment on the singularity structure of
non-perturbative thresholds. Singularities in thresholds are due to
states
becoming massless. In our case, such singularities occur whenever a
given term in the Eisenstein series diverges.
This happens for specific values of the
moduli.
However, it is important to note that
all these singularities can be  mapped
into the perturbative region
$\tau_2\to\infty$.
To put it otherwise, when the moduli are taken inside the fundamental
domain
of the $U$-duality group, then singularities occur at the boundaries.
This is equivalent to stating that the $U$-duality is not broken
by non-perturbative effects. This should be contrasted with cases
where
singularities
appear inside the moduli space, and break the original duality
group to a subgroup.

\vskip 5mm

\noindent 
{\bf Acknowledgements}. E. K. was  supported in part by EEC contract
TMR-ERBFMRXCT96-0090. We are grateful to  M. Henningson and N. Obers
for helpful discussions.

\vskip .5cm
\noindent
{\it Note added (1998). } Shortly after this article was released, a proof
of the conjectured ten-dimensional $R^4$ coupling Eq. \ref{a4} was
given from heterotic-type II duality in four dimensions 
\cite{apt}, and the perturbative non-renormalization theorem implied
by this result was demonstrated from $N=2, D=8$ superspace techniques 
\cite{berko}.

\vskip .5cm
\noindent
{\it Note added (jan. 2010).} The matrix $M_6$ in (5.27) and subsequent equation (5.28)
have been corrected, in agreement with Eq. (28) of \cite{piokir2}. Moreover, in order to produce an eigenmode of the Laplacian, the summation
in (5.28) should be restricted to null vectors with $n^i n_i=0$, as pointed out in \cite{Obers:1999um}.

\eject
\renewcommand{\theequation}{A.\arabic{equation}}
\centerline{\bf\large Appendix A: Expansion and Regularization of the
$SL(3)$}
\centerline{\bf\large Eisenstein series}
\setcounter{equation}{0}
\vskip .5cm

In this Appendix we give the expansion of the $SL(3)$ Eisenstein
series $E_s$ for arbitrary s. This is useful in order to derive the
regularized form $\hat E_{3\over 2}$.
The definition is
\be
E_s={\pi^s\over \Gamma(s)}\int_0^{\infty}{dt\over t^{1+s}}
{}~\hat{\sum_{m^i\in \Zints}}\exp\left[-{\pi\over t}
\left(\sum_{i,j=1}^3m^iM^{ij}m^j\right)\right]
\label{A1}\ee
$$=\n^{-s/3}{\pi^s\over \Gamma(s)}\int_0^{\infty}{dt\over t^{1+s}}
{}~\hat{\sum_{m^i\in \Zints}}\exp\left[-{\pi\over t} \left({m_3^2\over
\n}+{|m_1+m_2\tau+m_3B|^2\over \tau_2}\right)\right]
$$
Going through the same steps as in Section 4 we derive the following
expansion:
\be
E_s=2\n^{-s/3}\tau_2^s\z(2s)+2\sqrt{\pi}T_2\left(\tau_2\n^{1/3}
\right)^{{3\over
2}-s}{\Gamma\left(s-{1\over 2}\right)\over \Gamma(s)}\zeta(2s-1)+
\label{A2}\ee
$$+2\pi\n^{{2s\over 3}-1}{\z(2s-2)\over s-1}+I^s_D+\hat{\sum_{p,q\in  
\Zints}}I^s_{p,q}
$$
where
\be
I^s_D=2{\sqrt{\tau_2}\over \n^{s/3}}{\pi^s\over
\Gamma(s)}\sum_{m,n\not=0}
\left|{m\over n}\right|^{s-{1\over 2}}e^{2\pi imn\tau_1}
K_{s-{1\over
2}}(2\pi|mn|\tau_2)
\label{A3}\ee
\be
I^s_{p,q}=2{\n^{(s-3)/6}\over \tau_2^{(s-1)/2}}{\pi^s\over \Gamma(s)}
\sum_{m\not=0}\left|{p+q\tau\over m}\right|^{s-1}e^{2\pi
im(qB_1-(p+q\tau_1)B_2/\tau_2)}K_{s-1}\left(2\pi |m|{
|p+q\tau|\over \sqrt{\n\tau_2}}\right)
\label{A4}\ee
where the $K$ Bessel function arises through its integral representation:
\be
\int_0^{\infty} {dx \over x^{1-\lambda}} e^{-b/x-cx}
=2 \left|{ b\over c }\right|^{\lambda/2} K_{\lambda}(2\sqrt{|bc|})
\sp
K_{\lambda}(x)=K_{-\lambda}(x)
\sp
K_{1/2}(x)=\sqrt{\p \over 2x} e^{-x}
\ee

As a function of $s$, $E_s$ has potential simple
poles at
$s=1/2,1,3/2$. Indeed, $\z(s)$ has a simple pole at $s=1$ and
$\Gamma(s)$ a simple pole at $s=0$:
\be
\z(1+\epsilon)={1\over \epsilon}+\gamma+{\cal O}(\epsilon)
\sp
\Gamma(\epsilon)=1/\epsilon -\gamma+{\cal O}(\epsilon)
\label{A5}\ee
where $\gamma=0.577215...$ is the Euler constant.
{}From this we find that the residues ${\cal R}$ of the simple poles of
$E_s$ are
\be
{\cal R}_{1/2}={\cal R}_{1}=0\;\;\;,\;\;\;{\cal R}_{3/2}=2\pi
\label{A6}\ee
so that $E_{1}$ and $E_{1/2}$ are actually well defined.
On the other hand, for $s=3/2$, using
Eq.\ (\ref{199})
we obtain
\be
\hat E_{{3\over 2}}=2\zeta(3){\tau_2^{3/2}\over \nu^{1/2}}+{2\pi^2\over
3}T_2
+4\pi\log\n^{1/3}+I^{3/2}_D+\hat{\sum_{p,q\in \Zints}}I^{3/2}_{p,q}
\label{A7}\ee
with
\be
I^{3/2}_{p,q}=4\sum_{m\not=0}{1\over |m|}\exp\left[-2\pi |m|T_2
|p+q\tau|+2\pi im(qB_1-(p+q\tau_1)B_2/\tau_2)\right]
\label{A8}\ee
Finally we can rewrite the above result using the $SL(2,\Zint)$
invariant form
$f_{10}(\tau,\bar\tau)$ introduced in Eq.\ (\ref{a5}) as
\be
\hat E_{{3\over 2}}={f_{10}(\tau,\bar\tau)\over \nu^{1/2}}
+2\pi\log\n^{1/3}+\hat{\sum_{p,q\in \Zints}}I^{3/2}_{p,q}
\label{A9}\ee

Since we are also interested in other divergent $SL(3,\Zint)$
Eisenstein series, we would
like to
regularize the sum in a generic way, namely by introducing a
dimensionful
parameter $\m$.
This can be done by inserting a regulating function in Eq.\ (\ref{A1}):
\be
E^{\mu}_s={\pi^s\over \Gamma(s)}\int_0^{\infty}{dt\over t^{1+s}}
{}~\hat{\sum_{m^i\in \Zints}}\exp\left[-{\pi\over t}
\left(\sum_{i,j=1}^3m^iM^{ij}m^j\right)\right]R_{\mu}(t)
\label{A10}\ee
$$={\pi^s\over \Gamma(s)}\int_0^{\infty}{dt\over t^{1+s}}
{}~\hat{\sum_{m^i\in \Zints}}\exp\left[-{\pi\n^{1/3}\over t}
\left({m_3^2\over
\n}+{|m_1+m_2\tau+m_3B|^2\over \tau_2}\right)\right]R_{\mu}(t)
$$

Going through the same steps of the calculation we obtain
$$
E_{s,\rm reg}=2{\pi^s\over \Gamma(s)}\left[\int_{0}^{\infty}
dt~t^{s-1}R_{\mu}(1/t)\sum_{m=1}^{\infty}e^{-\pi t\n^{1/3}m^2/\tau_2}
+{\sqrt{\tau_2}\over \n^{1/6}}
\int_{0}^{\infty}
dt~t^{s-3/2}R_{\mu}(1/t)\sum_{m=1}^{\infty}e^{-\pi t\n^{1/3}\tau_2 m^2}
\right.
$$
$$\left.+\n^{-1/3}\int_{0}^{\infty}
dt~t^{s-2}R_{\mu}(1/t)\sum_{m=1}^{\infty}e^{-\pi t\n^{-2/3} m^2}\right]
+I_D^s+\hat{\sum_{p,q\in \Zints}}I^{s}_{p,q}+\cdots
$$
where the dots stand for setting the regulating function to 1 inside
the finite contributions $I^s_{D}$ and $I^s_{p,q}$.
For s=3/2 we can choose
\be
R_{\mu}=1-e^{-{\pi\over \mu^2 t}}
\label{A11}\ee
to obtain
\be
E^{\mu}_{3\over 2}=-{1\over 2}\log\mu^2+\gamma-\log 2+{1\over
3}\log\n+{f_{10}(\tau,\bar\tau)\over \nu^{1/2}}+\hat{\sum_{p,q\in
Z}}I^{3/2}_{p,q}+{\cal O}(\mu^2)
\label{A13}\ee
where we have used
\be
\sum_{m=1}^{\infty}\left[{1\over m}-{1\over \sqrt{m^2+\kappa^2}}\right]
=\gamma+\log(\kappa/2)+{\cal O}(1/\kappa)
\label{A14}\ee
Finally, a third way of regularizing the series, which is useful
for power-divergent series, is to subtract ``by hand'' the divergent piece.
In particular, we can define the order-1 Eisenstein series as
\be
E_{1,\rm reg}=\pi\int_0^{\infty} \left(  {1\over t^{2}}
{}~\hat{\sum_{m^i\in \Zints}}\exp\left[-{\pi\over t}
\left(\sum_{i,j=1}^3m^iM^{ij}m^j\right)\right]-
{1\over \sqrt{t}}\right) dt\ .
\label{A50}\ee
This way of regulating is motivated from perturbative threshold
corrections, and manifestly preserves the $SL(3,\Zint)$ symmetry.


\vskip 2cm

\begin{thebibliography}{99}

\bibitem{BK} 
  C.~Bachas and E.~Kiritsis,
  Nucl.\ Phys.\ Proc.\ Suppl.\  {\bf 55B} (1997) 194
  [arXiv:hep-th/9611205].

\bibitem{BFKOV} 
  C.~Bachas, C.~Fabre, E.~Kiritsis, N.~A.~Obers and P.~Vanhove,
  Nucl.\ Phys.\  B {\bf 509} (1998) 33
  [arXiv:hep-th/9707126].


\bibitem{gg} M.B. Green and M. Gutperle,  Phys. Lett. {\bf B398}
(1997) 69,
hep-th/9612127; hep-th/9701093.


\bibitem{vg2} 
  M.~B.~Green, M.~Gutperle and P.~Vanhove,
  Phys.\ Lett.\  B {\bf 409} (1997) 177
  [arXiv:hep-th/9706175].

\bibitem{sch} J.H. Schwarz, Nucl. Phys. {\bf B226} (1983) 269;\\
P.S. Howe and  P.C. West, Nucl. Phys. {\bf B238} (1984) 181.


\bibitem{sl} J.H. Schwarz,  Phys. Lett. {\bf B360} (1995) 13,
hep-th/9508143;
Phys. Lett. {\bf B367} (1996) 97,
hep-th/9510086;\\
E. Witten, Nucl. Phys. {\bf B460} (1996) 335, hep-th/9510135.

\bibitem{Roo}
E. Bergshoeff and M. de Roo, Nucl. Phys {\bf B328} (1989 439 \\
M. de Roo, H. Suelmann and A. Wiedmann, Phys. Lett {\bf B280} (1992)
39;
Nucl. Phys. {\bf B405} (1993) 326, hep-th/9210099

\bibitem{gri} M. Grisaru, A. Van de Ven and D. Zanon, Nucl. Phys. {\bf
B277}
(1986) 388, 409;\\
D. Gross and E. Witten, Nucl. Phys. {\bf B277} (1986) 1.



\bibitem{vw} C. Vafa and E. Witten, Nucl. Phys. {\bf B447} (1995) 261,
hep-th/9505053.


\bibitem{gsw} M.B. Green, J.H. Schwarz and E. Witten, {\it Superstring
Theory}
(Cambridge U.P,1985) Appendix 9.A

\bibitem{tsey}
A.A. Tseytlin, Nucl.Phys.{\bf B467} (1996) 383,
hep-th/9512081

\bibitem{julia} B. Julia, in the proceedings of the Nuffield Gravity
Workshop, Cambridge, June 22 - July 12, 1980.


\bibitem{ht}  C.M. Hull and  P.K. Townsend,  Nucl. Phys. {\bf B438}
(1995) 109,
hep-th/9410167.



\bibitem{sixguys}  
A.~Gregori, E.~Kiritsis, C.~Kounnas, N.~A.~Obers, P.~M.~Petropoulos and B.~Pioline,
  Nucl.\ Phys.\  B {\bf 510} (1998) 423
  [arXiv:hep-th/9708062].

\bibitem{kk} E. Kiritsis and C. Kounnas, Nucl. Phys. {\bf B442}
(1995) 472,  hep-th/9501020;\\
Nucl. Phys. {\bf B41} [Proc. Sup.] (1995) 331,  hep-th/9410212.

\bibitem{dkl} L. Dixon, V. Kaplunovsky and J. Louis, Nucl. Phys. {\bf
B355}
(1991) 649.


\bibitem{vg1}  M.~B.~Green and P.~Vanhove,
  Phys.\ Lett.\  B {\bf 408} (1997) 122
  [arXiv:hep-th/9704145].

\bibitem{pol} J. Polchinski, S. Chaudhuri and  C. Johnson,
hep-th/9602052;\\
J. Polchinski, hep-th/9611050.

\bibitem{ghm} M.B. Green, J. Harvey and G. Moore,
Class. Quant. Grav. {\bf 14} (1997) 47, hep-th/9605053.

\bibitem{pq} E. Witten, Nucl. Phys. {\bf B460}(1996)335, hep-th/9510135.

\bibitem{apt}   I.~Antoniadis, B.~Pioline and T.~R.~Taylor,
  Nucl.\ Phys.\  B {\bf 512} (1998) 61
  [arXiv:hep-th/9707222].

\bibitem{berko}   N.~Berkovits,
  Nucl.\ Phys.\  B {\bf 514} (1998) 191
  [arXiv:hep-th/9709116].



\bibitem{piokir2}  B.~Pioline and E.~Kiritsis,
  Phys.\ Lett.\  B {\bf 418} (1998) 61
  [arXiv:hep-th/9710078].

\bibitem{Obers:1999um}
  N.~A.~Obers and B.~Pioline,
  Commun.\ Math.\ Phys.\  {\bf 209} (2000) 275
  [arXiv:hep-th/9903113].


\end{thebibliography}
\end{document}